\documentclass{appolb}
\usepackage{epsfig}
\usepackage{pstricks}

\begin{document}
\title{Crossover transition to quark matter in heavy hybrid stars
}
\author{
David Edwin Alvarez Castillo
\address{Bogoliubov Laboratory for Theoretical Physics, JINR Dubna, 
Dubna, Russia\\ 
Instituto de Fisica, Universidad Autonoma de San Luis Potosi, S.L.P.,
Mexico}
\\[5mm]
Sanjin Beni\'{c}
\address{Physics Department, Faculty of Science, University of Zagreb, 
Zagreb, Croatia}
\\[5mm]
David Blaschke
\address{Institut Fizyki Teoretycznej, Uniwersytet Wroc{\l}awski, Wroc{\l}aw, 
Poland\\
Bogoliubov Laboratory for Theoretical Physics, JINR Dubna, Dubna, Russia}
\\[5mm]
Rafa{\l} {\L}astowiecki
\address{Institut Fizyki Teoretycznej, Uniwersytet Wroc\l awski, Wroc\l aw, 
Poland}
}
\maketitle
\begin{abstract}
We study the possibility that the transition from hadron matter 
to quark matter at vanishing temperatures
proceeds via crossover, similar to the crossover behavior
found with lattice QCD studies at high temperatures.
The purpose is to examine astrophysical consequences of this postulate
by constructing hybrid star sequences fulfilling current
experimental data.
\end{abstract}
  
\section{Introduction} 

Recent simulations of quantum chromodynamics (QCD) on the lattice 
\cite{Aoki:2006we,Bazavov:2011nk}
show that the hadron-to-quark matter transition in the
region of small quark chemical potential ($\mu\simeq 0$) in
the QCD phase diagram is a crossover.
To what extent this result persists in the regime for cold ($T=0$), 
dense ($\mu_B=3\mu > m_N$) matter is an open question eventually to be 
answered by heavy-ion collision 
experiments of the third generation such as NICA and FAIR.

An alternative is to measure masses and radii of compact stars,
where the Bayesian analysis is used to
invert the Tolman-Oppenheimer-Volkoff (TOV) equations \cite{Steiner:2012xt} 
and obtain the most probable equation of state (EoS) corresponding 
to a chosen set of observational constraints.
The challenge within this approach is a reliable measurement of neutron 
star radii. We quote results from millisecond pulsar timing analyses 
\cite{Bogdanov:2012md} rather than burst sources used in \cite{Steiner:2012xt}.

Prompted by recent findings of a second $2M_\odot$ neutron star 
\cite{Demorest:2010bx,Antoniadis:2013pzd} we reexamine the EoS for dense 
matter by constructing non-rotating sequences and studying the possibility 
of a hybrid star with $2M_\odot$.

The hybrid EoS is constructed from a non-local Nambu--Jona-Lasinio 
model (nl-NJL) \cite{Contrera:2010kz,Hell:2011ic,Benic:2013eqa} with
appreciable vector interaction strength \cite{Contrera:2012wj}, while
for the nuclear matter we use the DD2 EoS \cite{Typel:1999yq,Typel:2009sy}.

In this work we abandon the standard Maxwell construction for a 
first order phase transition, anticipating instead a crossover transition 
described by an interpolation of the pressure $p(\mu_B)$ as a  
thermodynamic potential for the above EoS. 
This procedure is equivalent to the one described in \cite{Masuda:2012ed} 
using energy density $\varepsilon$ versus baryon density $n_B$, which corrects
an earlier suggested inappropriate construction in the $p(n_B)$ plane
\cite{Masuda:2012kf}.
The construction leads to a characteristic stiffening on the hadron-dominated 
side followed by a softening and smooth joining to the quark-dominated side of 
the EoS.
While the former maybe due to quark substructure effects (Pauli blocking) 
initiating the hadron dissociation (Mott effect), 
the appearance of finite size structures (pasta phases) at the quark-hadron 
interface \cite{Yasutake:2012dw} and strong hadronic fluctuations 
\cite{Herbst:2010rf} might be responsible for the latter.

\section{Equation of state}

The thermodynamic potential for quark matter is provided by the nl-NJL model
\begin{equation}
\label{Omega}
\Omega = \Omega_\mathrm{cond} + \Omega_\mathrm{kin}^\mathrm{reg}
+\Omega_\mathrm{reg}^\mathrm{free}~,
\end{equation}
\begin{equation}
\Omega_\mathrm{cond} = 
\frac{\sigma_1^2+\kappa_\mathbf{p}^2\sigma_2^2
+\kappa_{p_4}^2\sigma_3^2}{2G_S}
-\frac{\omega^2}{2G_V}~,
\label{eq:omcond}
\end{equation}
\begin{equation}
\Omega_\mathrm{kin}^\mathrm{reg} = 
-2N_f N_c\int\frac{d^4 p}{(2\pi)^4}\log\left[\frac{\mathbf{p}^2A^2(\tilde{p}^2)
+ \tilde{p}_4^2C^2(\tilde{p}^2)
+ B^2(\tilde{p}^2)}{\tilde{p}^2 
+ m^2}\right]~,
\end{equation}
\begin{equation}
\Omega_\mathrm{reg}^\mathrm{free} = 
- \frac{N_c}{24\pi^2}\left[2\tilde{\mu}^3 \tilde{p}_F
- 5m^2 \tilde{\mu} \tilde{p}_F
+ 3m^4 \log\left(\frac{\tilde{p}_F+\tilde{\mu}}{m}\right)\right]~.
\end{equation}
Here $A(p^2) = 1+\sigma_2 f(p^2)$, $B(p^2)=1+\sigma_1 g(p^2)$ and
$C(p^2) = 1+\sigma_3 f(p^2)$, where $f(p^2)$ and $g(p^2)$ are appropriately 
chosen formfactors \cite{Contrera:2010kz}. 
We denote 
$\tilde{p} = (\mathbf{p},\tilde{p}_4)$, $\tilde{p}_4 = p_4-i\tilde{\mu}$.
The vector channel is introduced as a background field, similar
to the way the Polyakov loop is introduced in NJL models; via renormalization 
of the quark chemical potential
$\tilde{\mu} = \mu-\omega$ and completed by a 
classical term in the thermodynamic potential (\ref{eq:omcond}).

The pressure corresponds to the thermodynamic potential in equilibrium 
by $p_Q(\mu) = -\Omega$, where the latter is found from Eq.~\ref{Omega}
for a given chemical potential as a minimum with respect 
to variations of the mean fields
\begin{equation}
\frac{\partial \Omega}{\partial (\sigma_1, \sigma_2, \sigma_3)} = 0~.
\end{equation}
The value of the vector meanfield $\omega$ is found from the constraint 
of a given baryon density $n_B=\partial p_Q/\partial \mu_B$, namely
$\omega = G_V \ n_B(\tilde{\mu}_B)~.$
 
For describing dense nuclear matter we choose the DD2 EoS \cite{Typel:1999yq,Typel:2009sy}.
The transition region is constructed by a Gaussian interpolation
\begin{equation}
p(\mu_B) = 
\left\{ \begin{array}{ll} 
p_H(\mu_B) \ , & \mu_B <  \bar{\mu} \\
\left[p_H(\mu_B)-p_Q(\mu_B)\right]e^{-(\mu_B-\bar{\mu})^2/\Gamma^2}+p_Q(\mu_B) \ , &  \mu_B > \bar{\mu} \end{array} 
\right.
\label{eq:eoscross}
\end{equation}
where $\bar{\mu}$ and $\Gamma$ are parameters 
controlling the onset and the width of the transition, respectively.
This approach is equivalent to the crossover construction in the 
$\epsilon-n_B$ plane \cite{Masuda:2012ed} where a $\tanh$ function was used, 
but corrects the inappropriate construction suggested in \cite{Masuda:2012kf}.
Note that in Refs.~\cite{Blaschke:2013rma,Blaschke:2013ana} 
the crossover construction was utilized for interpolating between quark matter 
EoS for two different values of the vector coupling thus mimicking its medium 
dependence.
There the transition from the hadronic to the quark phase is seen as a
sharp first order, while here we assume a smooth crossover.

\begin{figure}
\centerline{
\includegraphics[width=0.45\textwidth]{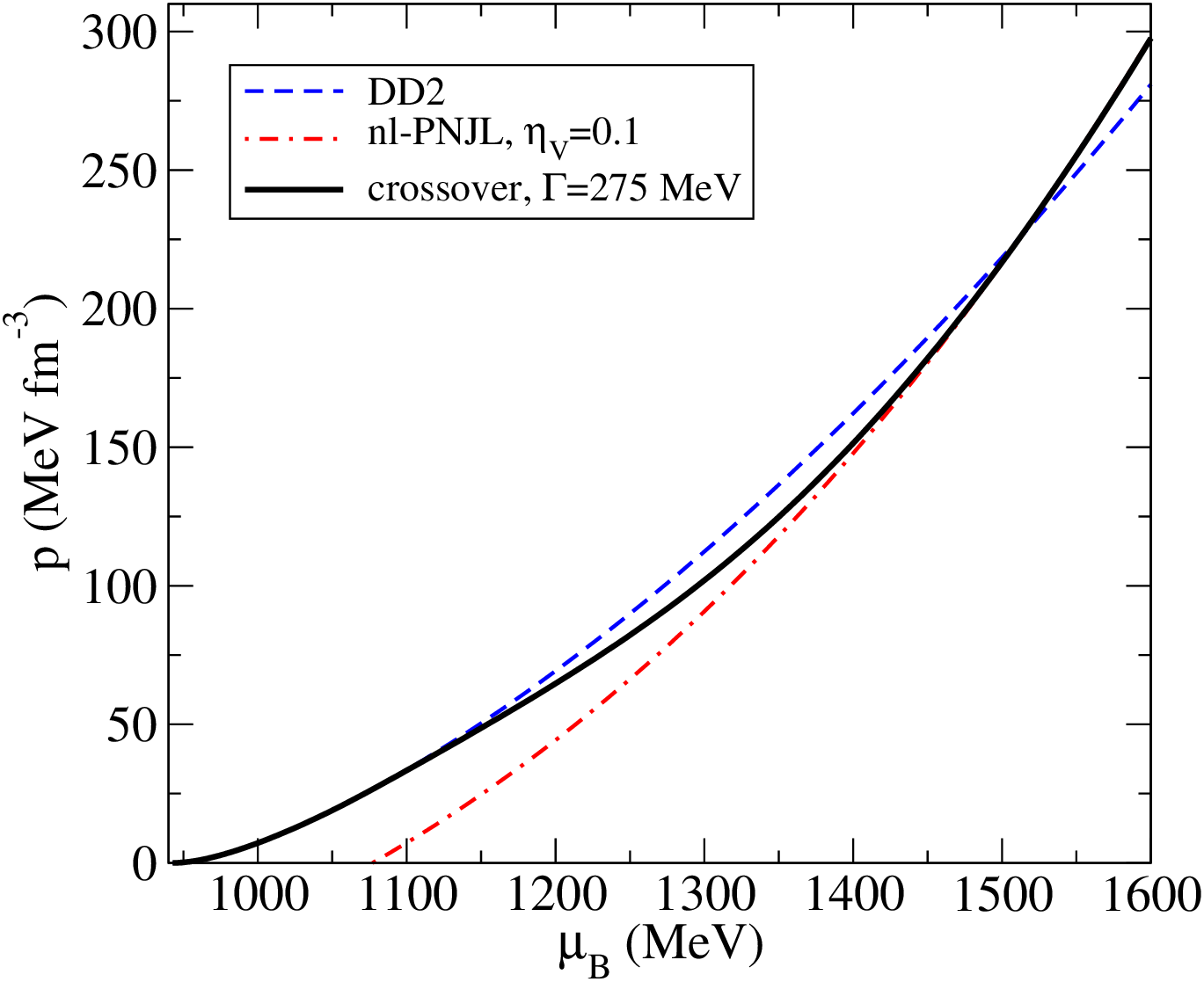}
\hspace{5mm}
\includegraphics[width=0.45\textwidth]{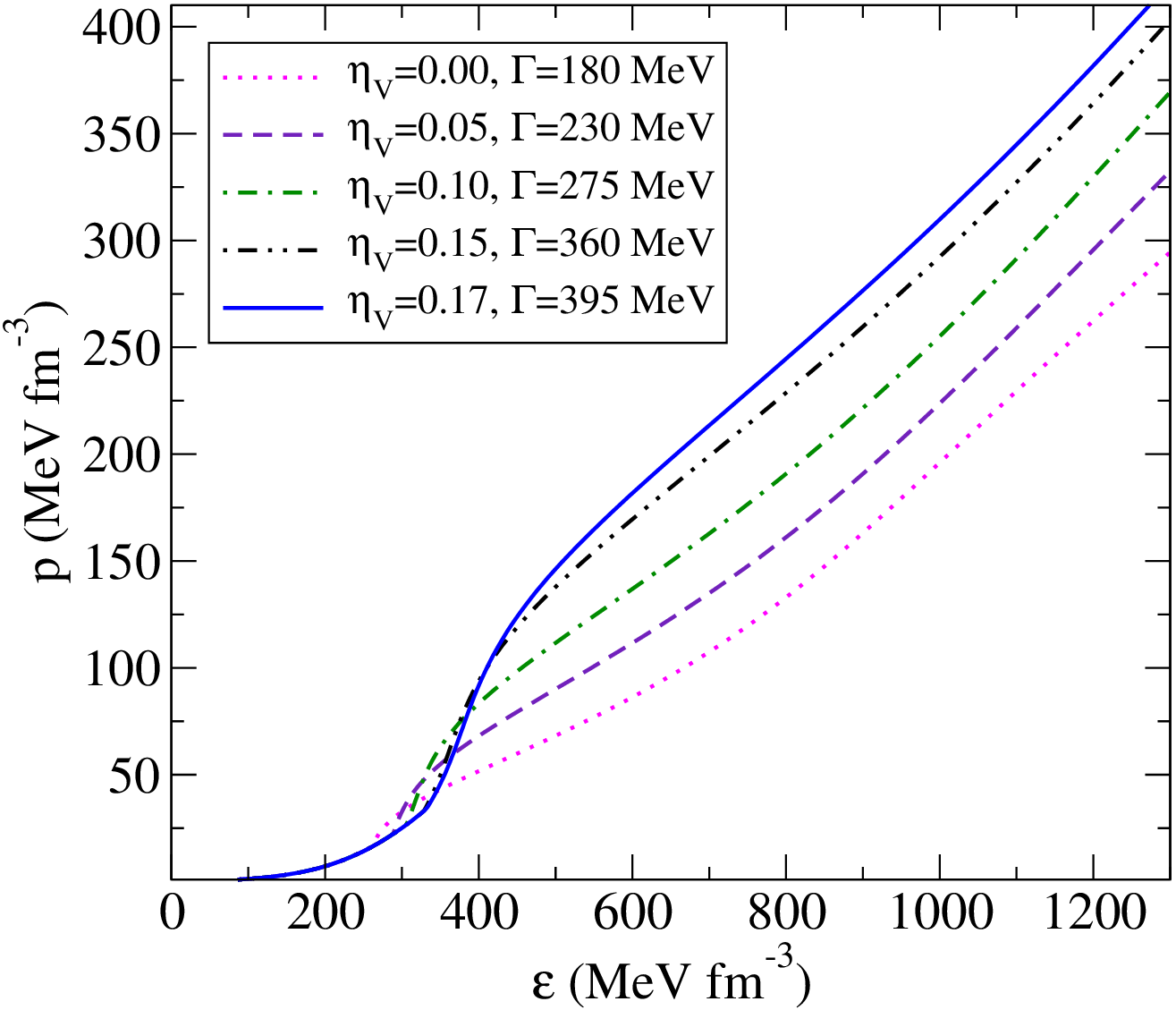}
}
\caption{Left panel shows the crossover construction Eq.~(\ref{eq:eoscross})
in the $p-\mu_B$ plane.
On the right panel we give the EoS used in this work.}
\label{fig:eos}
\end{figure}
There is an obvious benefit from our approach. 
With the Maxwell construction $p$ as a function of energy density $\epsilon$ 
in the transition region is flat, making the EoS soft, while
the crossover construction leads to a stiffening in the 
transition region.

We set $\bar{\mu}=\mu_c$, where $\mu_c$ is the onset of quark
matter in the nl-NJL model and use the 
minimal possible $\Gamma$ consistent with causality.
This leaves $\eta_V = G_V/G_S$ as a free parameter.
The resulting EoS are shown in the right panel of Fig.~\ref{fig:eos}.

\section{Astrophysical implications}

\begin{figure}
\centerline{
\includegraphics[width=0.45\textwidth]{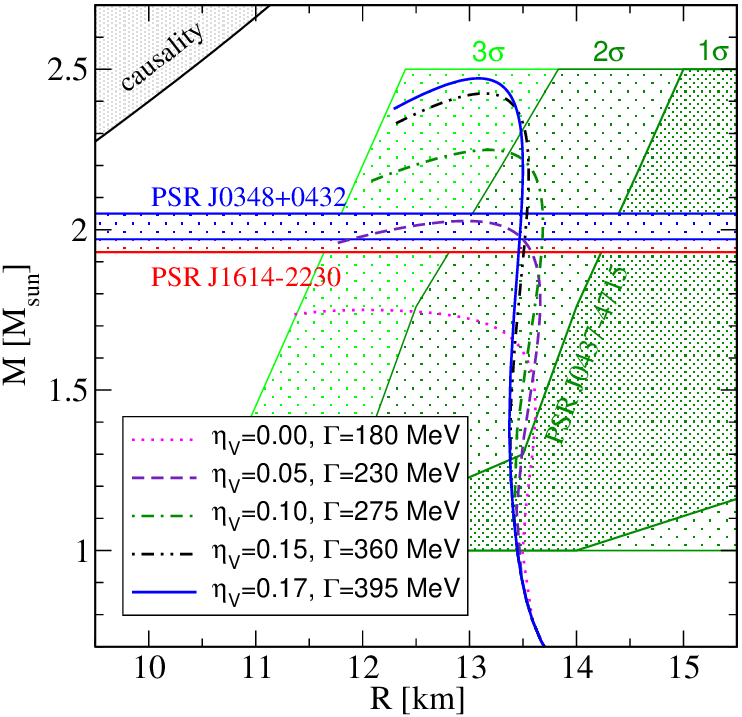}
\hspace{5mm}
\includegraphics[width=0.45\textwidth]{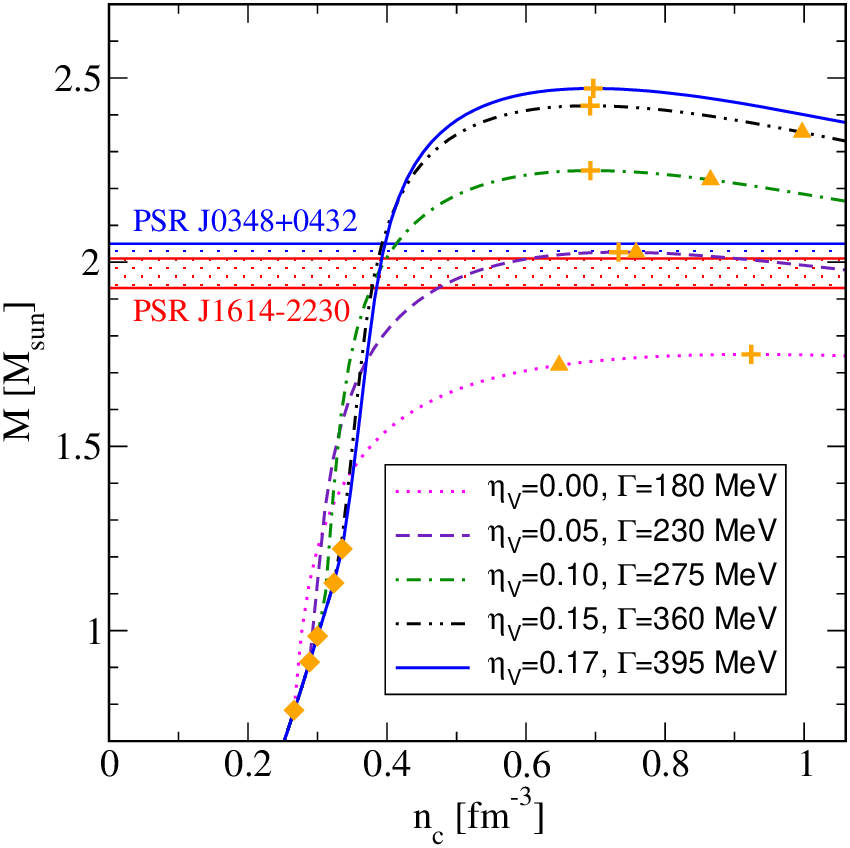}
}
\caption{The left panel shows sequences in the $M-R$ plane, while the right
panel gives mass as a function of central density for these sequences.
The orange diamonds represent the onset of the crossover region, 
while orange triangles denote its end corresponding to the onset of pure 
quark matter in the core of the neutron star. 
The orange plusses are the maximum mass configurations for the given EoS.
}
\label{fig:tov}
\end{figure}

It is known that within the Maxwell construction scheme hybrid stars with small
or almost zero vector coupling do not reach $2M_\odot$ before turning 
unstable \cite{Klahn:2013kga}.
The softness of quark matter is then regulated by a strong vector coupling 
channel.
However, the delay the quark matter onset caused by large vector couplings can 
be compensated by a strong diquark coupling.
In total, if quark matter appears through a first order transition, model 
calculations indicate that stable stars may require large couplings in both 
vector and diquark channels.

In this work we limit ourselves to the region of small vector coupling 
($\eta_V<0.17$) and offer an alternative mechanism that compensates such 
relative softness of the quark EoS.
We solve the TOV equations by using the EoS with the
crossover construction (\ref{eq:eoscross}).
The results shown in Fig.~\ref{fig:tov} are able to predict stable
stars reaching and exceeding $2M_\odot$ already with a small $\eta_V$.

Our calculations show that quark matter appears for sequences with mass 
heavier than $M\sim M_\odot$, and central densities higher than 
$n_c \sim 2n_0$ where $n_0 = 0.16$ fm$^{-3}$.
For vector couplings $\eta_V >0.05$ sequences
with masses above $2M_\odot$ do not have pure quark matter in their 
cores.
If we vaguely consider the crossover region as a mixed phase of hadrons
and quarks, we might describe these stars to have a hadronic mantle and a 
mixed phase core.
For mild vector coupling $\eta_V = 0.05$ the sequence with pure quark matter 
lies on the verge of stability, as shown by the triangle on the dashed line 
in the right panel of Fig.~\ref{fig:tov}.
In addition, it is interesting to note that this sequence lies within the 
$1~\sigma$ band of both the $2M_\odot$ constraint for PSR J0348-0432
\cite{Antoniadis:2013pzd} and PSR J1614-2230 \cite{Demorest:2010bx}. 

The possibility of having only a mixed phase in the $M>2M_\odot$ stars 
is due to the tension between the vector coupling and the width of the 
crossover.
With lower values of the vector coupling hadronic matter and quark matter EoS
lie closer to each other in the $p-\mu_B$ plane so that a smaller crossover 
region is needed to achieve a causal EoS.
By increasing the vector coupling the onset density of quark matter increases
and the two EoS separate, requiring a larger crossover region.

\section{Conclusions}

Requiring the transition from hadron to quark matter being unique, any viable
hybrid EoS model should be able to fulfill the constraint of the recent 
observational lower limit of $2M_\odot$  for the 
maximum mass of the corresponding hybrid star sequence.
For a wide class of NJL models this is not possible with the standard
Maxwell construction unless quark matter is in a color superconducting state
and has a strongly repulsive vector meanfield.

We have offered an alternative based on the requirement that the transition 
between quark and hadron matter is a crossover.
We have found that hybrid star configurations reaching or even exceeding 
the $2 M_\odot$ mass constraint have cores comprised of a mixed phase 
of quarks and hadrons.
This conclusion is similar to the one drawn when promoting the local charge 
neutrality condition to a global one via Gibbs construction 
\cite{Orsaria:2012je}.

Our work might be considered as a first step towards a microscopically based
construction of the transition from hadron to quark matter either via
pasta phases or beyond mean-field studies taking into account 
the quark substructure of hadrons and their dissociation in the 
dense medium \cite{Wergieluk:2012gd,Blaschke:2013zaa}.

\subsection*{Acknowledgements}
We are grateful for exciting discussions on
the subject of crossover EoS constructions 
to T. Fischer, H. Grigorian, P. Haensel, T. Hatsuda, M. Hempel,
T. Kl\"ahn, J. Lattimer, T. Maruyama, K. Masuda, I. Mishustin, 
J. Schaffner-Bielich, S. Typel, D. N. Voskresensky and N. Yasutake; 
and on neutron star radius measurements to M. C. Miller and J. Tr\"umper.  
This work was supported 
in part by 
the COST Action MP1304 ``NewCompStar'' and 
by NCN 
within the ``Maestro'' programme 
under contract No. DEC-2011/02/A/ST2/00306. 
D.A-C. received funding by the Bogoliubov-Infeld programme
for his collaboration with the University of Wroclaw.
S.B. was supported by the Ministry of Science, Education and Sports of Croatia 
through contract No. 119-0982930-1016.
D.B. acknowledges support by 
the Russian Fund for Basic Research 
under grant No. 11-02-01538-a and by EMMI for his participation in the Rapid 
Reaction Task Force Meeting ``Quark Matter in Compact Stars'' at FIAS 
Frankfurt.


\begin{thebibliography}{10}

\bibitem{Aoki:2006we}
  Y.~Aoki, G.~Endrodi, Z.~Fodor, S.~D.~Katz and K.~K.~Szabo,
  Nature {\bf 443} (2006) 675.

\bibitem{Bazavov:2011nk}
  A.~Bazavov {\it et al.},
  Phys.\ Rev.\ D {\bf 85} (2012) 054503.
  
\bibitem{Steiner:2012xt}
  A.~W.~Steiner, J.~M.~Lattimer and E.~F.~Brown,
  Astrophys.\ J.\  {\bf 765} (2013) L5.
  
\bibitem{Bogdanov:2012md}
  S.~Bogdanov,
  Astrophys.\ J.\  {\bf 762} (2013) 96.
  [arXiv:1211.6113 [astro-ph.HE]].
  
\bibitem{Demorest:2010bx}
  P.~Demorest {\it et al.},
  Nature {\bf 467} (2010) 1081.
  
\bibitem{Antoniadis:2013pzd}
  J.~Antoniadis {\it et al.},
  Science {\bf 340} (2013) 6131.

\bibitem{Contrera:2010kz}
  G.A.~Contrera, M.~Orsaria and N.N.~Scoccola,
  Phys.\ Rev.\ D {\bf 82} (2010) 054026.

\bibitem{Hell:2011ic}
  T.~Hell, K.~Kashiwa and W.~Weise,
  Phys.\ Rev.\ D {\bf 83} (2011) 114008.
  
\bibitem{Benic:2013eqa}
  S.~Benic, D.~Blaschke, G.A.~Contrera and D.~Horvatic,
  arXiv:1306.0588.
  
\bibitem{Contrera:2012wj}
  G.A.~Contrera, A.G.~Grunfeld and D.~Blaschke,
  arXiv:1207.4890 [hep-ph].
  
\bibitem{Typel:1999yq}
  S.~Typel and H.~H.~Wolter,
  Nucl.\ Phys.\ A {\bf 656} (1999) 331.
  
\bibitem{Typel:2009sy}
  S.~Typel, G.~Ropke, T.~Klahn, D.~Blaschke and H.~H.~Wolter,
  Phys.\ Rev.\ C {\bf 81}, 015803 (2010)
  
\bibitem{Masuda:2012ed}
  K.~Masuda, T.~Hatsuda and T.~Takatsuka,
  PTEP {\bf 2013} (2013) 7,  073D01.
  
\bibitem{Masuda:2012kf}
  K.~Masuda, T.~Hatsuda and T.~Takatsuka,
  Astrophys.\ J.\  {\bf 764} (2013) 12.

\bibitem{Yasutake:2012dw}
  N.~Yasutake {\it et al.},
  arXiv:1208.0427 [astro-ph.HE].
  
\bibitem{Herbst:2010rf}
  T.K.~Herbst, J.M.~Pawlowski and B.-J.~Schaefer,
  Phys. Lett. B {\bf 696} (2011) 58.
  
\bibitem{Blaschke:2013rma}
 D.~Blaschke {\it et al.},
  PoS ConfinementX {\bf } (2012) 249;
  [arXiv:1302.6275 [hep-ph]].
  
\bibitem{Blaschke:2013ana}
  D.~Blaschke, D.~E.~Alvarez-Castillo and S.~Benic,
  PoS CPOD2013 (2013) 063.  
  
\bibitem{Klahn:2013kga}
  T.~Kl\"ahn, D.~Blaschke and R.~{\L}astowiecki,
  Phys.\ Rev.\ D {\bf 88} (2013) 085001.
  
\bibitem{Orsaria:2012je}
  M.~Orsaria, H.~Rodrigues, F.~Weber and G.~A.~Contrera,
  Phys.\ Rev.\ D {\bf 87} (2013) 023001.
  
\bibitem{Wergieluk:2012gd}
  A.~Wergieluk, D.~Blaschke, Yu.L.~Kalinovsky and A.~Friesen,
  Phys. Part. Nucl. Lett. {\bf 7} (2013) in press; [arXiv:1212.5245 [nucl-th]].

\bibitem{Blaschke:2013zaa}
  D.~Blaschke, D.~Zablocki, M.~Buballa and G.~R\"opke,
  arXiv:1305.3907 [hep-ph].

\end{thebibliography}
\end{document}